\documentclass[prd,showpacs,twocolumn,
amsmath,amssymb,superscriptaddress,floatfix,nofootinbib]{revtex4-1}
\usepackage[utf8]{inputenc}
\usepackage[sort&compress]{natbib}
\usepackage[normalem]{ulem}
\usepackage{lipsum} 
\usepackage{bm}
\usepackage{times}
\usepackage{amssymb,amsbsy,amsmath,amsfonts}
\usepackage{graphicx}
\usepackage{float}
\usepackage{color}
\usepackage{morefloats}
\usepackage{rotating}
\usepackage{srcltx}
\usepackage{slashed}
\usepackage{subfigure}
\usepackage{multirow}
\usepackage{verbatim}
\usepackage{hyperref}
\usepackage{tabularx}
\usepackage{adjustbox}
\usepackage[dvipsnames]{xcolor}
\usepackage{nicefrac}
\renewcommand{\arraystretch}{1.2}

\usepackage{hyperref}
\hypersetup{
	colorlinks	=true,
	urlcolor	=blue,
	linkcolor	=blue,
	citecolor	=blue,
	pdftitle	={},
	pdfauthor	={Zhi-Wei Liu, Jun-Xu Lu, Ming-Zhu Liu, Li-Sheng Geng},
	pdfsubject	={$\Sigma_c\bar{D}^{(*)}$ correlation function}
	}

\begin{document}

\title{
Probing the di-$J/\Psi$ interaction and the nature of $X(6200)$ with femtoscopic correlation functions
}

\author{Zhi-Wei Liu}
\affiliation{School of Physics, Beihang University, Beijing 102206, China}

\author{Jia-Ming Xie}
\affiliation{School of Physics, Beihang University, Beijing 102206, China}

\author{Jun-Xu Lu}
\affiliation{School of Physics, Beihang University, Beijing 102206, China}

\author{Li-Sheng Geng}
\email[Corresponding author: ]{lisheng.geng@buaa.edu.cn}
\affiliation{School of Physics, Beihang University, Beijing 102206, China}
\affiliation{Sino-French Carbon Neutrality Research Center, \'Ecole Centrale de P\'ekin/School of General Engineering, Beihang University, Beijing 100191, China}
\affiliation{Peng Huanwu Collaborative Center for Research and Education, Beihang University, Beijing 100191, China}
\affiliation{Beijing Key Laboratory of Advanced Nuclear Materials and Physics, Beihang University, Beijing 102206, China }
\affiliation{Southern Center for Nuclear-Science Theory (SCNT), Institute of Modern Physics, Chinese Academy of Sciences, Huizhou 516000, China}

\begin{abstract}
Recent re-analyses of the di-$J/\Psi$ invariant mass spectra reveal a state near the di-$J/\Psi$ threshold, referred to as the $X(6200)$. Yet the nature of this near‑threshold pole-- whether it is a resonant, bound, or virtual state -- remains unresolved due to our limited understanding of the di-$J/\Psi$ interaction. To address this question, we predict the di-$J/\Psi$ and $J/\Psi\Psi(2S)$ femtoscopic correlation functions based on the Koonin-Pratt formula with a Gaussian source and the coupled‑channel dynamics. Our results show that the di-$J/\Psi$ correlation function exhibits distinctly different behaviors in each scenario, especially for small source sizes ($ R\sim1$ fm), providing a clear experimental observable to distinguish the nature of $X(6200)$. These distinguishing features persist even when quantum statistical effects and coupled-channel dynamics are included and show negligible sensitivity to off-shell ambiguities. Given the high $J/\Psi$ production rates and clean detection channels at the LHC, we hope that these discoveries will stimulate further experimental studies and help clarify the nature of double-vector-charmonium interactions and the nonperturbative dynamics of fully-heavy tetraquark systems.
\end{abstract}


\maketitle

\section{Introduction}
As the fundamental theory of the strong interaction, Quantum Chromodynamics (QCD) governs the dynamics of quarks and gluons. In the low-energy regime, QCD requires quarks to be confined within color-singlets, i.e., hadrons. According to the conventional quark model, hadrons can be classified into $q\bar{q}$ mesons and $qqq$ baryons~\cite{Gell-Mann:1964ewy,Zweig:1964ruk}. However, since 2003, numerous $qq\bar{q}\bar{q}$ tetraquark and $qqq\bar{q}\bar{q}$ pentaquark states beyond the above two configurations have been discovered experimentally~\cite{Belle:2003nnu,BaBar:2005hhc,BESIII:2013ris,LHCb:2015yax,LHCb:2021vvq,Olsen:2017bmm}, marking the dawn of a new era in hadron physics (see, e.g.,
Refs.~\cite{Oset:2016lyh,Richard:2016eis,Hosaka:2016ypm,Chen:2016qju,Esposito:2016noz,Lebed:2016hpi,Guo:2017jvc,Ali:2017jda,Karliner:2017qhf,Liu:2019zoy,vanBeveren:2020eis,Chen:2022asf,Mai:2022eur,Liu:2024uxn} for recent reviews). In particular, fully-heavy tetraquark states, which contain no light quarks, provide a unique platform to gain further insight into the color confinement.

In recent years, experimental studies of all-charm tetraquarks have advanced considerably. 
In 2020, the LHCb Collaboration reported a narrow structure, denoted as $X(6900)$, together with a broad enhancement in the near double-$J/\Psi$ threshold region from 6.2 to 6.8 GeV in the di-$J/\Psi$ invariant-mass spectrum~\cite{LHCb:2020bwg}. 
The ATLAS and CMS Collaborations subsequently confirmed the $X(6900)$ and observed additional broad structures in the di-$J/\Psi$ and $J/\Psi\Psi(2S)$ spectra~\cite{ATLAS:2023bft,CMS:2023owd}. 
More recently, ATLAS and CMS have also measured the ratio of partial decay widths $\Gamma_{X(6900)\rightarrow J/\Psi+\Psi(2S)}/\Gamma_{X(6900)\rightarrow \mathrm{di}\text{-}J/\Psi}$ and determined the quantum numbers of the $X(6900)$ to be $J^{PC}=2^{++}$~\cite{ATLAS:2025nsd,CMS:2025fpt}. 
Taken together, the existence and basic properties of the $X(6900)$ appear to be well established by these three experiments, whereas the interpretation of the broader accompanying enhancement still requires additional data.

It is worth noting that the sizable mass gap between the $X(6900)$ and the di-$J/\Psi$ threshold exceeds a typical radial or orbital excitation energy, hinting at the possible existence of an all-charm tetraquark below the mass of the $X(6900)$. 
Recent re-analyses of the aforementioned invariant-mass spectra have revealed strong evidence for a pole near the di-$J/\Psi$ threshold, commonly referred to as the $X(6200)$~\cite{Dong:2020nwy,Liang:2021fzr,Wang:2022jmb,Niu:2022jqp,Huang:2024jin,Song:2024ykq}. 
Furthermore, it has been shown that soft-gluon exchanges between two $J/\Psi$ mesons, which hadronize into two pions at long distances, can bind the di-$J/\Psi$ system and generate a near-threshold pole~\cite{Dong:2021lkh,Nefediev:2021pww}. 
However, substantial ambiguity remains concerning whether the $X(6200)$ corresponds to a di-$J/\Psi$ resonant, virtual, or bound state. 
Even within the same coupled-channel frameworks, all three scenarios can reproduce the invariant-mass spectra with comparable quality~\cite{Dong:2020nwy,Liang:2021fzr,Song:2024ykq}. 
There is little doubt that this ambiguity largely stems from our still limited understanding of the di-$J/\Psi$ interaction.

In the last decade, by measuring momentum correlation functions, femtoscopy has provided valuable insights into rare hadron-hadron scattering processes~\cite{STAR:2014dcy,STAR:2015kha,STAR:2025jwe,ALICE:2019gcn,ALICE:2019hdt,ALICE:2020mfd,ALICE:2021cpv,Fabbietti:2020bfg,Zhang:2025tfd}, which also triggered a large number of theoretical studies~\cite{Morita:2014kza,Haidenbauer:2018jvl,Kamiya:2019uiw,Liu:2022nec,Molina:2023oeu,Yan:2024aap,Achenbach:2024wgy,Li:2024tvo,Ge:2025put,Ikeno:2025kwe,Liu:2025eqw,Xie:2025xew}, especially in the charm and bottom sectors~\cite{Kamiya:2022thy,Liu:2023uly,Liu:2023wfo,Vidana:2023olz,Ikeno:2023ojl,Torres-Rincon:2023qll,Feijoo:2023sfe,Khemchandani:2023xup,Li:2024tof,Albaladejo:2024lam,
Liu:2024nac,Liu:2025rci,Liu:2024uxn,Liu:2025oar,Liu:2025wwx,Shen:2025qpj}. It is worthwhile noting that the ALICE and STAR Collaborations have recently successfully measured the $\bar{D}N$, $D^{(*)}\pi$, and $D^{(*)}K$ CFs~\cite{ALICE:2022enj,ALICE:2024bhk,RoyChowdhury:2024vff,Zhang:2025szg}, paving the way for precision studies of the strong interaction in the charm sector. Our previous works demonstrated the significant potential of femtoscopic correlation functions for identifying near-threshold exotic states. By employing the $D^0D^{*-}/D^0D_s^{*-}$ correlation functions, Ref.~\cite{Liu:2024nac} demonstrates a means to distinguish between resonant, virtual, and bound states for the $Z_c(3900)/Z_{cs}(3985)$, overcoming the general difficulty caused by phase-space suppression, which renders such an identification difficult when relying solely on invariant mass spectra. Subsequently, Ref.~\cite{Liu:2025wwx} demonstrated that the $D_s^+D_s^-$ correlation function remains effective even in the presence of the Coulomb interaction for $X(3960)$. 

Motivated by the success of recent experiments and theoretical developments, this work aims (i) to predict the di-$J/\Psi$ correlation function as a new observable that can motivate future measurements of the di-$J/\Psi$ interaction, and (ii) to demonstrate that this correlation function can help elucidate the nature of the near-threshold $X(6200)$ pole, even when quantum-statistical and coupled-channel effects are taken into account. The remaining part of this paper is organized as follows. Sect. II reviews the di-$J/\Psi$ interaction in the coupled-channel framework and the formalism for calculating correlation functions. Sect. III presents predictions for the di-$J/\Psi$ and $J/\Psi\Psi(2S)$ correlation functions. Conclusions and outlook are given in Sect. IV.

\section{Theoretical Framework}
In this section, we briefly recall the formalism of double-vector-charmonium  interactions within the coupled-channel framework and explain how to calculate the femtoscopic correlation function for two identical particles. Following Ref.~\cite{Dong:2020nwy}, the $S$-wave potentials for two- and three-channel models with the configurations $J/\Psi J/\Psi-J/\Psi\Psi(2S)$ and $J/\Psi J/\Psi-J/\Psi\Psi(2S)-J/\Psi\Psi(3770)$, are expressed as symmetric matrices,
\begin{subequations}
\begin{align}
  V_{\rm 2ch}(\sqrt{s})&=
  \left(
  \begin{array}{cc}
    a_1+b_1k_1^2 & c \\
    & a_2+b_2k_2^2
  \end{array}
  \right),\label{Eq:V2ch}\\
  \nonumber\\
  V_{\rm 3ch}(\sqrt{s})&=
  \left(
  \begin{array}{ccc}
    a_{11} & a_{12} & a_{13} \\
     & a_{22} & a_{23} \\
     &  & a_{33}
  \end{array}
  \right)\label{Eq:V3ch},
\end{align}
\end{subequations}
where $\sqrt{s}$ is the center-of-mass (c.m.) energy, $i$th channel c.m. momentum $k_i=\frac{\sqrt{s-(m_{i1}+m_{i2})^2}\sqrt{s-(m_{i1}-m_{i2})^2}}{2\sqrt{s}}$, $m_{i1}$ and $m_{i2}$ represent the particle masses in the $i$th channel, and \{$a_1$, $a_2$, $b_1$, $b_2$, $c$\} and \{$a_{11}$, $a_{12}$, $a_{13}$, $a_{22}$, $a_{23}$, $a_{33}$\} are real low-energy constants (LECs) to be determined. Since the above potentials are separable, the scattering $T$-matrix can be obtained as
\begin{align}
  T(\sqrt{s})=\left[\mathbb{I}-V_{\rm 2(3)ch}(\sqrt{s})\cdot G(\sqrt{s})\right]^{-1}V_{\rm 2(3)ch}(\sqrt{s}).\label{Eq:LS}
\end{align}
Here, $G$ is a diagonal matrix for the intermediate two-body propagators, defined by the dimensionally regularized two-point scalar loop function,
\begin{align}
  G_i(\sqrt{s})&=\frac{1}{16\pi^2}\left\{a(\mu)+\log\left(\frac{m_{i1}^2}{\mu^2}\right)+\frac{m_{i2}^2-m_{i1}^2+s}{2s}\right.\nonumber\\
  &\times\log\left(\frac{m_{i2}^2}{m_{i1}^2}\right)+\frac{k}{\sqrt{s}}\left[\log\left(\frac{2k_i\sqrt{s}+s+\Delta_i}{2k_i\sqrt{s}-s+\Delta_i}\right)\right.\nonumber\\
  &\left.\left.+\log\left(\frac{2k_i\sqrt{s}+s-\Delta_i}{2k_i\sqrt{s}-s-\Delta_i}\right)\right]\right\},\label{Eq:LF}
\end{align}
where the dimensional regularization scale $\mu$ and the subtraction constant $a(\mu)$ are set at $1$ GeV and $-3$~\cite{Dong:2020nwy}, respectively, and $\Delta_i=m_{i1}^2-m_{i2}^2$.

According to the Koonin‑Pratt formula~\cite{Koonin:1977fh,Pratt:1990zq,Bauer:1992ffu}, the femtoscopic correlation function is determined by two fundamental components: 1) the particle-emitting source $S_{12}$ produced in relativistic $pp$, $pA$, and $AA$ collisions; 2) the scattering wave function $\psi$ of the relative motion of the particle pair under study. For the particle-emitting source $S_{12}$, numerous methods have been developed to eliminate its ambiguity. A prominent example is the resonance source model, a data-driven approach proposed by the ALICE Collaboration, which has been widely adopted in femtoscopic studies~\cite{ALICE:2020mfd,ALICE:2021cpv,ALICE:2022enj}. In this model, the source is characterized by a Gaussian core that emits all primordial particles and exhibits a clear transverse mass ($m_{\rm T}$) scaling~\cite{ALICE:2020ibs,ALICE:2023sjd}, complemented by an exponential tail arising from strongly decaying resonances. Other recent advancements include the optical deblurring algorithm for imaging sources in heavy-ion collisions~\cite{Xu:2024dnd}, reconstruction of proton-emitting sources from experimental correlation functions using deep neural networks within an automatic differentiation framework~\cite{Wang:2024bpl}, simulations of single-particle phase-space distributions from transport models~\cite{Wang:2024yke}, and reconstructing source functions via the Tikhonov regularization~\cite{Xiong:2025bmd}. In the present work, given expected minimal contributions from the strong decays feeding into charm mesons~\cite{ALICE:2022enj}, adopting a Gaussian source with a single parameter $R$, namely, $S_{12}(r)=\exp\left[-r^2/(4R^2)\right]/(2\sqrt{\pi}R)^3$, can be a very good starting point in this exploratory study. The source size $R$ can be anchored to the proton-proton correlation data by measuring $m_{\rm T}$ of the pair of interest in future experiments.

Another essential component -- the relative wave function $\psi$ -- encodes the final‑state interaction of interest. In the present work, since the potentials in Eqs.~\eqref{Eq:V2ch} and \eqref{Eq:V3ch} are formulated in momentum space, it is advantageous first to compute the reaction amplitude $T$‑matrix by solving Eq.~\eqref{Eq:LS}, and then derive the relative wave function via the relation $|\psi\rangle = |\varphi\rangle + G T |\varphi\rangle$, where $|\varphi\rangle$ is the free wave function. More specifically, the relative wave function $\psi_{ij}$ for non-identical particles has the following form~\cite{Ohnishi:2016elb,ExHIC:2017smd}:
\begin{align}
  \psi_{ij}(\boldsymbol{r})=\delta_{ij}\left[e^{{\rm i}\boldsymbol{k}\cdot\boldsymbol{r}}-j_0(kr)\right]+\psi_{ij}^0(r),\label{Eq:WF}
\end{align}
where $j_0(kr)$ is the spherical Bessel function, and the subscripts $i$ and $j$ denote the outgoing and incoming channel, respectively. The $S$-wave relative wave function $\psi_{ij}^0$ is written as~\cite{Haidenbauer:2018jvl,Liu:2023uly,Vidana:2023olz}
\begin{align}
  \psi_{ij}^0(r)&=\delta_{ij}j_0(kr)+\int_0^\infty\frac{{\rm d^3}k'}{(2\pi)^3}\frac{\omega_{i1}(k')+\omega_{i2}(k')}{2\omega_{i1}(k')\omega_{i2}(k')}\nonumber\\
  &\times\frac{T_{ij}(\sqrt{s})\theta(q_{\rm max}-k)\theta(q_{\rm max}-k')\cdot j_0(k'r)}{s-\left[\omega_{i1}(k')+\omega_{i2}(k')\right]^2+{\rm i}\varepsilon},\label{Eq:SWF}
\end{align}
where $\omega_i(k')=\sqrt{m_i^2+k^{\prime2}}$. Notably, the above expression incorporates the factor $\theta(q_{\rm max}-k)\theta(q_{\rm max}-k')$ to specify the off-shell behavior of the separable $T$-matrix. In general, the factor $\theta(q_{\rm max}-k)$ is inoperative since one always works with values of the momenta smaller than the sharp cutoff $q_{\rm amx}$. The factor $\theta(q_{\rm max}-k')$, which controls the off-shell behavior, will be absorbed as the upper limit of the integral. Substituting Eqs.~\eqref{Eq:WF} and \eqref{Eq:SWF} into the Koonin‑Pratt formula, one can express the correlation function for non-identical particles as
\begin{align}
  C(k)&=1+\int_0^\infty4\pi r^2{\rm d}r~S_{12}(r)~\bigg[\sum_i\nu_i\big|\delta_{ij}j_0(kr)\nonumber\\
  &+T_{ij}(\sqrt{s})\widetilde{G}_i(r,\sqrt{s})\big|^2-\left|j_0(kr)\right|^2\bigg],\label{Eq:CF1}
\end{align}
where $\nu_i$ is the weight for each component of the multichannel wave function, and the sum runs over all possible coupled channels. For simplicity, we assume that the weights are the same and equal to $1$ (hereafter the same). The quantity $\widetilde{G}$ is given by~\cite{Vidana:2023olz}
\begin{align}
  \widetilde{G}_i(r,\sqrt{s})&=\int_0^{|\boldsymbol{q}|<q_{\rm max}}\frac{{\rm d^3}k'}{(2\pi)^3}\frac{\omega_{i1}(k')+\omega_{i2}(k')}{2\omega_{i1}(k')\omega_{i2}(k')}\nonumber\\
  &\times\frac{j_0(k'r)}{s-\left[\omega_{i1}(k')+\omega_{i2}(k')\right]^2+{\rm i}\varepsilon}.\label{Eq:G}
\end{align}

For systems of two identical particles, quantum statistical effects have to be taken into account, which requires that the relative wave function should be symmetrized or antisymmetrized with respect to the exchange of the coordinates of two particles (namely, $\boldsymbol{r}\rightarrow\boldsymbol{-r}$)~\cite{ExHIC:2017smd,Liu:2022nec}. For the di-$J/\Psi$ system (similar to the di-deuteron system~\cite{Ge:2025put}), the relative wave function can be decomposed into a symmetric spatial component with even parity (spin-singlet and spin-quintet~\footnote{Since Ref.~\cite{Dong:2020nwy} does not distinguish between the spin-singlet and spin-quintet di-$J/\Psi$ interactions, this work likewise does not differentiate between the corresponding two wave functions.}) and an antisymmetric spatial component with odd parity (spin-triplet), namely,
\begin{subequations}
\begin{align}
  \psi_{ij,E}(\boldsymbol{r})&=\frac{1}{\sqrt{2}}\left[\psi_{ij}(\boldsymbol{r})+\psi_{ij}(-\boldsymbol{r})\right]\nonumber\\
  &=\sqrt{2}\left[\delta_{ij}\cos(\boldsymbol{k}\cdot\boldsymbol{r})-\delta_{ij}j_0(kr)+\psi_{ij}^0(r)\right],\label{Eq:EWF}\\
  \psi_{ij,O}(\boldsymbol{r})&=\frac{1}{\sqrt{2}}\left[\psi_{ij}(\boldsymbol{r})-\psi_{ij}(-\boldsymbol{r})\right]\nonumber\\
  &=\sqrt{2}{\rm i}\left[\delta_{ij}\sin(\boldsymbol{k}\cdot\boldsymbol{r})\right].\label{Eq:OWF}
\end{align}
\end{subequations}
Considering that the spin dependence is not resolved in the standard measurements of correlation functions, the di-$J/\Psi$ correlation function should be averaged over the spin states. Substituting Eqs.~\eqref{Eq:EWF}, \eqref{Eq:OWF}, and \eqref{Eq:SWF} into the Koonin‑Pratt formula and taking the spin average, one can express the di-$J/\Psi$ correlation function as
\begin{align}
  C(k)&=1+\frac{1}{3}e^{-4k^2R^2}+\frac{4}{3}\int_0^\infty4\pi r^2{\rm d}r~S_{12}(r)\nonumber\\
  &\times\left[\sum_i\nu_i\left|\delta_{ij}j_0(kr)+T_{ij}(\sqrt{s})\widetilde{G}_i(r,\sqrt{s})\right|^2\right.\nonumber\\
  &-\left|j_0(kr)\right|^2\bigg],\label{Eq:CF2}
\end{align}
where the second term $\frac{1}{3}e^{-4k^2R^2}$ arises from the quantum statistical effect (so-called Bose-Einstein correlation), which enhances the correlation due to the symmetrization of the wave function for spin-1 bosons.

\section{Results and discussions}

As mentioned above, $\{a_1, a_2, b_1, b_2, c\}$ and $\{a_{11}, a_{12}, a_{13}, a_{22}, a_{23}, a_{33}\}$ denote the unknown parameters describing the double-vector-charmonium interactions in the two- and three-channel models, respectively. These parameters can be determined by fitting to the di-$J/\Psi$ invariant-mass spectrum. In Ref.~\cite{Dong:2020nwy}, the authors carried out such fits to the LHCb data, obtaining one parameter set for the two-channel model and two sets for the three-channel model. All three models successfully reproduce the experimental spectrum (see Figs.~1 and 2 of Ref.~\cite{Dong:2020nwy}) and predict the robust presence of a near-threshold state, $X(6200)$. More recently, the same authors updated their analysis by simultaneously incorporating data from the LHCb, ATLAS, and CMS Collaborations, reaching conclusions consistent with their earlier study~\cite{Song:2024ykq}. 
In the present work, we adopt the fitted results of Ref.~\cite{Dong:2020nwy} as inputs\footnote{The parameters appearing in Eqs.~\eqref{Eq:V2ch} and \eqref{Eq:V3ch} are obtained from those listed in Ref.~\cite{Dong:2020nwy} by multiplying them by the factor $\prod_i^4 \sqrt{2m_i}$.}. Before presenting the correlation functions, we briefly summarize the pole properties associated with these three interactions. As shown in Table~\ref{Tab:pole}, the two-channel interaction yields a resonance pole, the three-channel interaction (Fit~1) produces a bound-state pole, and the three-channel interaction (Fit~2) gives a virtual-state pole, all of which are consistent with the findings of Ref.~\cite{Dong:2020nwy}.

\begin{table}[htbp]
\centering
\setlength{\tabcolsep}{6.8pt}
\renewcommand{\arraystretch}{1.8}
\caption{Properties of the near-threshold $X(6200)$ predicted by the two- and three-channel models~\cite{Dong:2020nwy}. The $J/\Psi J/\Psi$, $J/\Psi\Psi(2S)$, and $J/\Psi\Psi(3770)$ thresholds are $6194$, $6783$, and $6870$ MeV, respectively.}\label{Tab:pole}
\begin{tabular}{cccc}
\hline
\hline
Model  & Riemann sheet  & Pole [MeV]  & Feature   \\
\hline
2-channel & $(-,+)$  & $6202-11i$  & resonant  \\
3-channel (Fit 1)  & $(+,+,+)$  & $6159$  & bound  \\
3-channel (Fit 2)  & $(-,+,+)$  & $6185$  & virtual  \\
\hline
\hline
\end{tabular}
\end{table}

\begin{figure*}[htbp]
  \centering
  \includegraphics[width=0.98\textwidth]{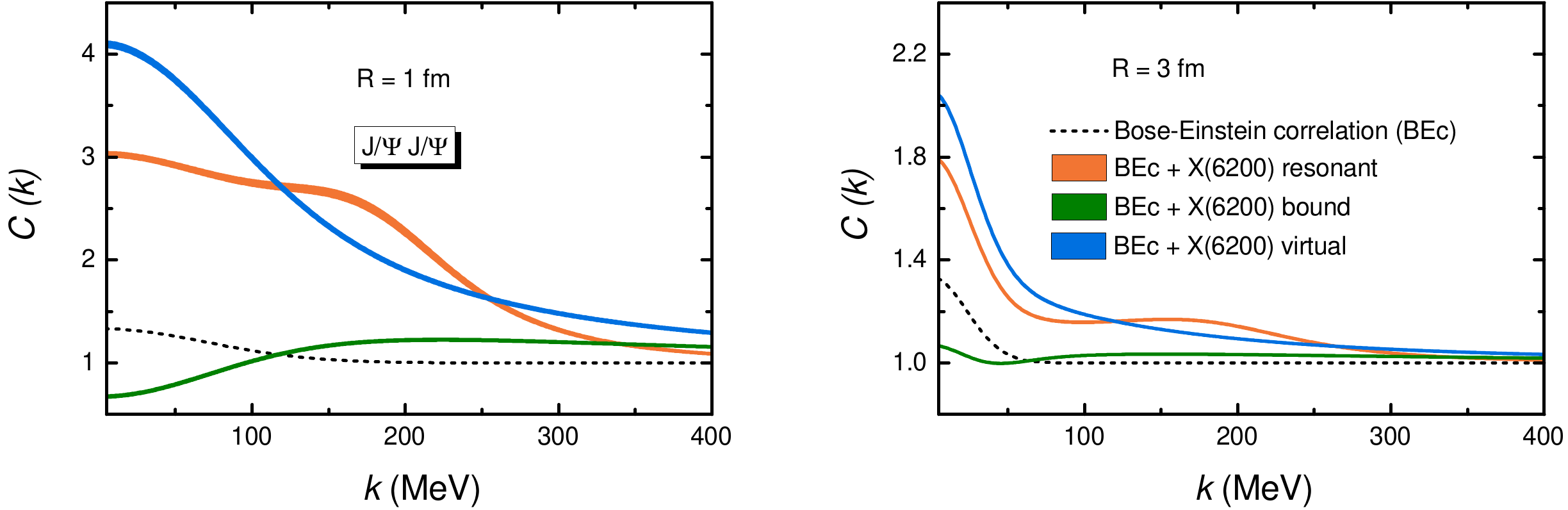}
  \caption{Di-$J/\Psi$ correlation function as a function of the relative momentum $k$ for different source sizes $R = 1$ and $3$ fm. The results are calculated with
  Eq.~\eqref{Eq:CF2} and three types of interactions. The bands reflect the variation in the sharp cutoff over the range $q_{\rm max}=800-1500$ MeV. The result obtained by considering only the quantum statistical effect (the so-called Bose-Einstein correlation) is also shown for comparison.}\label{Fig:JJ}
\end{figure*}

Figure~\ref{Fig:JJ} shows the di-$J/\Psi$ correlation function as a function of the relative momentum for source sizes $R=1$ and $3$~fm, under three distinct interaction models: the two-channel model and the two three-channel models (Fit~1 and Fit~2). The purely Bose-Einstein correlation (dashed curve) is also displayed, highlighting how final-state interactions substantially modify the baseline quantum-statistical enhancement. As expected, the correlation functions exhibit markedly different behaviors depending on the nature of the near-threshold pole. 
For $R=1$~fm, the bound-state scenario (three-channel, Fit~1) shows a pronounced suppression in the low-momentum region. In contrast, the virtual-state scenario (three-channel, Fit~2) exhibits a clear enhancement across the entire momentum range. The resonance scenario (two-channel model) yields a moderate enhancement at low momentum, followed by a rapid decrease at momenta slightly above the resonant energy. As discussed in Ref.~\cite{Liu:2024nac}, these distinct patterns are closely correlated with the corresponding scattering phase shifts~$\delta$: the bound-state case has $\delta$ starting at $180^\circ$ and gradually decreasing; the virtual-state case has $\delta$ starting at $0^\circ$ and increasing with momentum; and the resonance case has $\delta$ starting at $0^\circ$, rising with momentum, and undergoing a rapid increase---crossing $90^\circ$---near the resonance energy.
Although the inclusion of coupled-channel effects complicates the quantitative calculation, the qualitative behavior of the correlation functions remains preserved in all three scenarios. When the source size increases to $R=3$~fm, the differences between the three cases become less pronounced, yet still remain distinguishable.

As discussed above, the differences in the phase shifts account for the distinct features of the correlation functions in the three scenarios. In addition, the opposite behaviors exhibited in the virtual-state and bound-state cases can be naturally traced to the opposite signs of their corresponding scattering lengths. To provide experimental support for these features, we use nucleon--nucleon correlation functions as an illustrative example. It is well known that the $^3S_1$--$^3D_1$ nucleon--nucleon interaction produces a bound state (the deuteron), whereas the $^1S_0$ interaction generates a virtual state.
Recent measurements of neutron--neutron correlation functions have revealed a significant enhancement over the entire momentum range, particularly at low momenta~\cite{Si:2025eou}---a characteristic signature of a virtual state. In contrast, neutron--proton correlation functions, measured as early as 2001, exhibit a suppression at intermediate to high momenta~\cite{Ghetti:2001gm}. Since the measured neutron--proton correlation function represents an average over the spin-singlet and spin-triplet components, subtracting the positive contribution from the spin-singlet channel (corresponding to the virtual-state scenario) leaves the observed suppression attributable entirely to the spin-triplet channel (corresponding to the bound-state scenario). This provides direct experimental evidence for the typical correlation-function behavior associated with a bound state.
From a physical perspective, this behavior is intuitive: pairs that form a bound state are removed from the correlation yield, resulting in a suppression of the correlation function.

\begin{figure*}[htbp]
  \centering
  \includegraphics[width=0.98\textwidth]{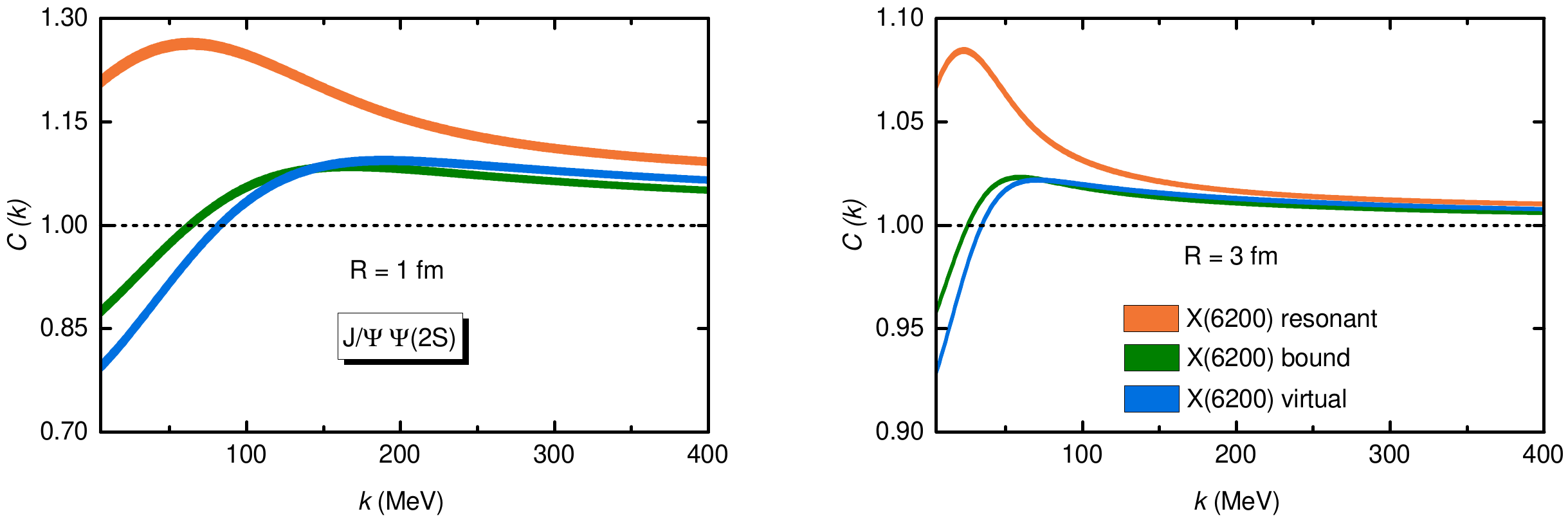}
  \caption{$J/\Psi\Psi(2S)$ correlation function as a function of the relative momentum $k$ for different source sizes $R = 1$ and $3$ fm. The results are calculated with Eq.~\eqref{Eq:CF1} and three types of interactions. The bands reflect the variation in the sharp cutoff over the range $q_{\rm max}=800-1500$ MeV.}\label{Fig:JJ2S}
\end{figure*}

To further distinguish the interactions among these three models, we present the $J/\Psi\Psi(2S)$ correlation function, as shown in Fig.~\ref{Fig:JJ2S}. Unlike the di-$J/\Psi$ case, here the correlation functions do not include the Bose-Einstein correlation term, as the particles are non-identical. The overall magnitude of the $J/\Psi\Psi(2S)$ correlation function is significantly smaller than that in Fig.~\ref{Fig:JJ}, reflecting the weaker interaction in this channel. Since the $J/\Psi\Psi(2S)$ threshold is significantly higher than the $X(6200)$ position, the $J/\Psi\Psi(2S)$ correlation functions no longer exhibit the characteristic features of resonance, virtual, or bound states. In addition, it is interesting to see that the di-$J/\Psi$ correlation functions predicted by the two-channel model and the three-channel model (Fit 2) exhibit slight overlap, while their predicted $J/\Psi\Psi(2S)$ correlation functions are distinctly separated.

Before closing our discussion, it is instructive to briefly comment on the off-shell ambiguity in strong interactions, which has been noted to alter the short-distance behavior of relative wave functions, thereby introducing theoretical uncertainties in calculations of femtoscopic correlation functions~\cite{Epelbaum:2025aan}. Nonetheless, as argued in the same work ~\cite{Epelbaum:2025aan}: \textit{``On the other hand, realistic models of hadronic interactions are constrained by physical principles like, e.g., pion-exchange dominance at large distances and often share similarities when it comes to modeling the short-distance behavior. The remaining off-shell ambiguities may therefore be expected to be less pronounced in practice.''} Further quantitative studies have demonstrated the relative mildness of these off-shell effects within realistic interaction models for the two-body problem~\cite{Gobel:2025afq,Molina:2025lzw}. In this work, we quantify the uncertainty associated with off-shell behavior by varying the sharp cutoff $q_{\rm max}$ from $800$ to $1500$ MeV. The bands in Figs.~\ref{Fig:JJ} and \ref{Fig:JJ2S} represent the resulting variation. It is seen that the influence of off-shell ambiguity on the di-$J/\Psi$ and $J/\Psi\Psi(2S)$ correlation functions is mild, almost negligible.

\section{Summary and outlook}

In this work, we have investigated the di-$J/\Psi$ and $J/\Psi\Psi(2S)$ femtoscopic correlation functions within a coupled-channel framework, considering three possible scenarios for the near-threshold $X(6200)$ state: resonant, bound, or virtual. Our results show that the di-$J/\Psi$ correlation function serves as a sensitive probe: for $R=1$ fm, the bound-state scenario leads to a strong suppression at low momentum, the virtual-state scenario produces a broad enhancement, and the resonance scenario exhibits a moderate enhancement followed by a rapid drop near the resonance energy. These distinctive patterns remain clearly observable even after incorporating quantum statistical symmetrization and coupled-channel effects, while the $J/\Psi\Psi(2S)$ correlation function provides complementary information with weaker sensitivity to the pole structure. The mild dependence on off-shell cutoffs further underscores the robustness of our predictions.

The abundant production of $J/\Psi$ mesons at the LHC~\cite{LHCb:2020bwg} and their clean reconstruction through the electromagnetic modes $e^+e^-$ and $\mu^+\mu^-$ (branching fractions $\sim$6\% each~\cite{ParticleDataGroup:2024cfk}) make them ideal probes for the di-$J/\Psi$ correlation function.
With the upgraded ALICE apparatus and the larger data sets expected in Runs 3-5~\cite{ALICE:2022wwr}, measurements of the di-$J/\Psi$ and $J/\Psi\Psi(2S)$ correlation functions appear feasible in the near future, which would provide valuable insight into double-vector-charmonium interactions and the nonperturbative dynamics of fully-heavy tetraquark systems.

\emph{Acknowledgments.} This work is partly supported by the National Science Foundation of China under Grant No. W2543006 and Nos. 12435007 and 1252200936, and the National Key R\&D Program of China under Grant No. 2023YFA1606703. Zhi-Wei Liu acknowledges support from the National Natural Science Foundation of China under Grant No.12405133, No.12347180, China Postdoctoral Science Foundation under Grant No.2023M740189, and the Postdoctoral Fellowship Program of CPSF under Grant No.GZC20233381.

\bibliography{diJPsi}

\clearpage
\end{document}